\documentclass[twocolumn,showpacs,preprintnumbers,amsmath,amssymb]{revtex4}
\usepackage{graphicx}
\usepackage{dcolumn}
\usepackage{bm}
\newcommand{\rr}{{\bf r}}

\newcommand{\vv}{{\bf v}}

\newcommand{\FF}{{\bf F}}

\newcommand{\nn}{{\bf n}}

\newcommand{\etal}{{\it et al.\ }}
\newcommand{\BE}{\begin{equation}}
\newcommand{\EE}{\end{equation}}
\newcommand{\be}{\begin{equation}}
\newcommand{\ee}{\end{equation}}
\begin{document}

\preprint{ }

\title{Pressure Driven Flow of Polymer Solutions in Nanoscale Slit Pores
\footnote{Submitted to J. Chem. Phys., Oct. 2006}} 
\author{Jaime A. Millan$^{1}$, Wenhua Jiang$^{2}$, Mohamed Laradji$^{1,3}$\footnote{
Electronic mail: mlaradji@memphis.edu} and Yongmei Wang$^{2}$}
\affiliation {$^1$Department of Physics, The University of Memphis, Memphis, TN 38152 \\
$^2$Department of Chemistry, The University of Memphis, Memphis, TN 38152 \\
$^3$MEMPHYS--Center for Biomembrane Physics, University of Southern Denmark, Odense, DK-5230, Denmark}
\date{\today}

\begin{abstract}

Polymer solutions subject to pressure driven flow and in nanoscale slit pores are 
systematically investigated using the dissipative particle dynamics approach.
We investigated the effect of molecular weight, polymer concentration and flow rate
on the profiles across the channel of the fluid and polymer velocities, polymers density, and the
three components of the polymers radius of gyration. 
We found that the mean streaming fluid velocity decreases as the polymer molecular weight 
or/and polymer concentration is increased, and that the deviation of the velocity profile 
from the parabolic profile is accentuated with increase 
in polymer molecular weight or concentration. We also found that the distribution
of polymers conformation is highly anisotropic and non-uniform across the channel.
The polymer density profile is also found to be non-uniform, exhibiting a local minimum
in the center-plane followed by two symmetric peaks. We found a migration of the polymer chains
either from or towards the walls. For relatively long chains, as compared to the thickness
of the slit, a migration towards the walls is observed. However, for relatively short chains,
a migration away from the walls is observed. 
\end{abstract}


\maketitle

\section{INTRODUCTION}

The transport of polymer solutions in small
confining geometries remains a long stranding topic with many unanswered questions. The understanding of the transport 
of polymer solutions is important to chromatography, electrophoresis, adhesion, lubrication, polymer processing,
oil recovery, etc. Recently, there has been an increasing interest in the areas of macromolecular transport in 
channels with widths comparable to size of a single  molecule~\cite{agarwal94}. 
For example, the understanding of the structure and dynamics of polymeric molecules in micro-channels 
is important to DNA sequencing in channels with widths ranging from $10$ to $50 \mu{\rm m}$~\cite{fang05},
DNA delivery through micro-capillaries in gene therapy, and to lab-on-chip applications that involve polymers~\cite{han00}. 
Issues of particular interest pertinent to polymer solutions in the presence of laminar flow is the mass 
distribution of polymers across the channel, the polymers conformational distribution, and the effect
of the polymer chains on the profile of the solution velocity field
~\cite{muller90,ausserre91,agarwal94,horn00,shrewsberry01,fan03,fang05,woo04_1,woo04_2,jendrejack04,ma05,usta05,usta06,khare06}.

Past experiments were performed to investigate the conformational changes of polymers 
in the presence of elongational flow by Perkins {\it et al.}~\cite{perkins97} and in
the presence of shear flow by Smith {\it et al.} ~\cite{smith99} through tracking fluorescently 
labeled DNA molecules. These experiments showed that the polymer chains become elongated along the flow direction.
Several experiments indicated the existence of a depletion layer in the polymers near the confining
walls~\cite{muller90,ausserre91,agarwal94,horn00,shrewsberry01}.
The recent experiment by Horn {\it et al.}~\cite{horn00} on a dilute solution of a high molecular weight 
polyisobutene in a very viscous fluid observed that the polymer solution exhibits flow with an apparent 
slip at the boundary. 

It is well known that at equilibrium conditions, polymer chains near walls, and in the non-adsorbed regime, 
are sterically depleted from the wall due their reduced conformational entropy~\cite{degennes79}. 
In the presence of a pressure-driven (Poiseuille) flow, the shear rate across the channel is non-uniform. 
As a result, one expects a higher concentration of polymers in the mid-section of
the channel since, there, the local shear is lowest, leading thereby to an apparent
slip at the walls~\cite{fang05}. Due to the additional stretching of the polymer chains near the walls,
as a result of the shear stresses originating from finite shear rate, thermodynamic arguments predict
that the thickness of the depletion layer is thinner when compared to that at equilibrium conditions. 
Consequently, the steric effect of the walls on the polymer chains is less severe than at equilibrium conditions.
However, experiments~\cite{agarwal94} on dilute polymer solutions 
demonstrated that even in the case of uniform shear flow, the polymer chains
migrate away from the wall leading to a depletion layer that is wider than at equilibrium. 
Therefore, the thermodynamic arguments alone are unable to fully account for the
migration of the polymer chains observed in experiments. It is important to note that
since these systems are not at equilibrium, thermodynamic arguments are not applicable. 

The understanding of the migration of the polymer chains away from the wall must therefore take into
account hydrodynamic interactions in the solution. However, such understanding is complicated by 
the non-Newtonian properties of polymer solutions. Theories have therefore been based on simplified systems
rather than on polymer solutions. Previous theoretical and computational studies that focused on the
density profiles of polymers in uniform shear flow ({\it i.e.}, Couette flow) or pressure
driven flow ({\it i.e.}, Poiseuille flow) have been conflicting~\cite{woo04_1,woo04_2,ma05,usta05,usta06}. 
Previous Brownian dynamics simulations showed a migration of the polymer 
chains towards the wall~\cite{woo04_1,woo04_2}. In these studies, however, wall-monomer hydrodynamic 
interactions are not accounted for.
A recent kinetic theory by Ma and Graham~\cite{ma05} of an infinitesimally dilute solution of 
dumbbells predicts a migration of the dumbbell particles away from the wall towards the mid-section 
of the channel. A key reason for the migration of the dumbbells away from the wall is attributed to
the hydrodynamic interaction between the wall and the dumbbell, which 
leads to a deterministic migration from the wall, while the effect of
the Brownian motion is to homogenize the solution and therefore to counter the effect of the  wall-monomer
hydrodynamic interaction. This cross-stream migration was recently investigated 
numerically by Jendrejack {\it et al.} using a Brownian dynamics simulation that
self-consistently accounts for both monomer-monomer hydrodynamic interaction and wall-monomer hydrodynamic
interaction~\cite{jendrejack04}. In this study, they observed that in the presence
of pressure driven flow, the polymer density profile exhibits a local minimum (a dip)
in the middle of the pore, which decreases as the flow rate is increased. They also observed that the
depletion region from the walls is amplified as the flow rate is increased. The depletion from the wall
becomes substantial for large flow rates.
In the more recent simulations by Usta {\it et al.}~\cite{usta05,usta06},
which are based on a bead model for the polymer chains coupled to a lattice Boltzmann 
description for the solvent, the polymer density profile was also found to exhibit a minimum 
in the center-plane of the slit followed 
by two symmetric maxima. Near the wall, a migration away from the wall is observed when 
the ratio between the thickness of the slit and the equilibrium radius of gyration of the polymer chains is large,
{\it i.e.}, in the weak confinement regime. However, when confinement is increased, a migration
towards the wall is observed. This result contrasts that of Jendrejack {\it et al.}~\cite{jendrejack04},
where even in the strong confinement regime, a migration away from the wall is observed. 
In the presence of uniform shear flow, Usta {\it et al.}~\cite{usta06} 
observed that the density profile does not exhibit a peak
in the mid-section of the slit, instead of a dip. The presence of a dip in the mid-section of the slit must therefore
be associated with gradients in the shear rate, present in pressure driven flow.

Very recently, Khare {\it et al.} used molecular dynamics to study the density profile of dilute polymer solutions
in both uniform shear flow and pressure driven flow~\cite{khare06}. In the case of pressure driven flow,
they also observed a local minimum in the polymer density profile. However, the depletion layer is only weakly
modified by flow.  As in Usta {\it et al.}'s work~\cite{usta06}, Kare {\it et al.} also observed that 
the polymer density profile peaks in the mid section of the slit in the case of uniform shear flow.
A recent simulation by Fan {\it et al.}~\cite{fan03} 
based on dissipative particle dynamics of dilute polymer solutions
in pressure driven flow in a slit channel focused on the velocity profile of the polymer solutions. 
In their study however, they studied the migration of dimers instead of chains, and they
observed very weak migration~\cite{fan03}. Therefore, the issue of migration clearly remains settled, warranting 
further investigation. 

A large number of studies during recent years have proved that dissipative particle dynamics (DPD)  
is a successful explicit computational approach in the investigation of both equilibrium and non-equilibrium 
properties of complex fluids. A recent study by Ripoll {\it et al.} showed, in particular, 
that DPD correctly describe large scale and mesoscopic hydrodynamics in fluids~\cite{ripoll01}.
Using extensive DPD simulations, we recently investigated equilibrium dynamics of polymer 
chains in the dilute and concentrated regimes, and both in bulk and in confining geometries. We found that 
the chains dynamics is well described 
by the Zimm model in the dilute regime, and as expected a crossover to Rouse dynamics is observed
as the polymer concentration is increased~\cite{jiang06,jiang07}. These studies validate the usefulness
of DPD in correctly describing the dynamics of polymer solutions. We therefore used the DPD
approach in studying the dynamics of polymer solutions in pressure driven flow.

In the present article, we present results based on DPD simulations of polymer solutions
in nanoscale slits and in the presence of pressure driven flow. Our study is 
performed on systems with various polymer volume fractions, chain length, and flow rate. We will particularly
discuss the velocity profile of the solution across the channel, the conformational distribution of the polymer chains
and their mass distribution. In Sec. 3, we present results of Poiseuille flow of a simple fluid using DPD.
In Sec. 4, the model and computational approach is presented. In Sec. 3, we present and
discuss the numerical results. Finally, we summarize and conclude in Sec. 5.

\section{MODEL AND METHOD}

In this work, we investigate the dynamics of dilute and semi-dilute polymer solutions under pressure
driven flow using the dissipative particle dynamics (DPD) approach. DPD was first
introduced by Hoogerbrugge and Koelman more than a decade ago~\cite{hoogerbrugge92}, and was cast
in its present form about five years later~\cite{espagnol95,espagnol97}.  DPD is reminiscent
of molecular dynamics, but is more appropriate for the investigation of generic properties of
macromolecular systems. The use of soft repulsive interactions in DPD, allows for larger
integration time increments than in a typical molecular dynamics simulation using
Lennard-Jones interactions. Thus time and length scales much larger than those in
atomistic molecular dynamics simulations can be probed by the DPD approach.
Furthermore, DPD uses pairwise random and dissipative forces between neighboring
particles, which are interrelated through the fluctuation-dissipation theorem.
The pairwise nature of these forces ensures the local conservation of momentum, a necessary
condition for correct long-range hydrodynamics~\cite{pagonabarraga98,ripoll01}.
The DPD approach was recently used for the study of flow of colloidal suspensions~\cite{palma06} and flow  
of dilute polymer solutions~\cite{fan03,fan06}.

\begin{figure}
\includegraphics[scale=0.37]{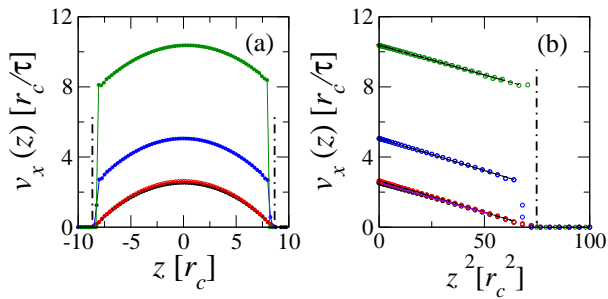}
\caption{(a) Velocity field along the $x$-axis vs. $z$ for the case of a driving force, $f_x=0.02\epsilon/r_c$.
Curves from bottom to top correspond to wall-solvent interaction parameter 
$a_{ws}=3\epsilon/r_c$, $10\epsilon/r_c$, $20\epsilon/r_c$,
and $30\epsilon/r_c$, respectively. (b) The velocity profile vs. $z^2$. Not that $v_x(z)$ is quadratic in $z$ regardless
of the wall-solvent interaction. The vertical dot-dashed lines indicate the positions of the walls.
Notice that curves for $a_{ws}=3\epsilon/r_c$ and $10\epsilon/r_c$ overlap with each other.}
\label{fig1}
\end{figure}

\begin{figure}
\includegraphics[scale=0.37]{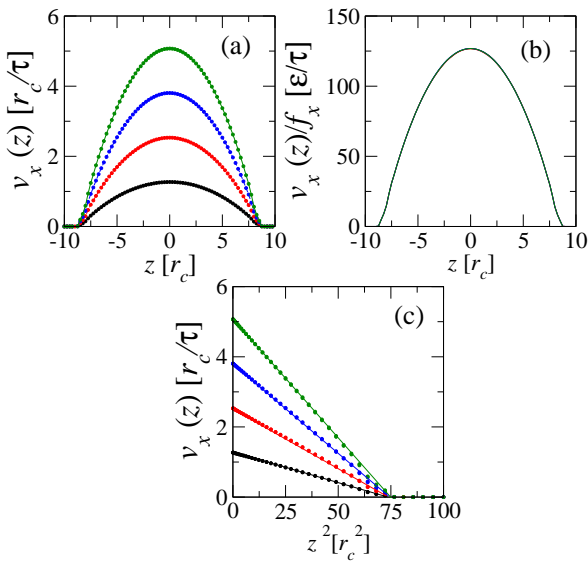}
\caption{(a) Velocity field along the $x$-axis vs. $z$ for $a_{ws}=3\epsilon/r_c$.
Curves from bottom to top correspond to a driving force $f_x=0.01$, 0.02, 0.03, and $0.04\epsilon/r_c$.
(b) The velocity profile normalized by the driving force vs. $z$. The excellent collapse in (b) 
implies that the velocity profile follows Eq.~(\ref{eq:poiseuille}). (c) The velocity profile 
vs. $z^2$.}
\label{fig2}
\end{figure}

In the DPD approach, two particles, $i$ and $j$, interact with each other via three pairwise forces corresponding
to a conservative force, $\FF^{(C)}_{ij}$, a dissipative force, $\FF^{(D)}_{ij}$, and a random force,
$\FF^{(R)}_{ij}$. These three forces are respectively given by,
\begin{eqnarray}
\FF^{(C)}_{ij}&=& a_{ij}\omega(r_{ij})\nn_{ij}, \label{eq:conservative-force}\\
\FF^{(D)}_{ij}&=& \gamma_{ij}\omega^2(r_{ij})(\hat\rr_{ij}\cdot\vv_{ij})\nn_{ij},
\label{eq:dissipative-force}\\
\FF^{(R)}_{ij}&=& \sigma_{ij}(\Delta t)^{1/2}\omega(r_{ij})\theta_{ij}\nn_{ij},
\label{eq:random-force}
\end{eqnarray}
where $\rr_{ij}=\rr_j-\rr_i$, ${\nn}_{ij}={\rr_{ij}}/{r_{ij}}$, and $\vv_{ij}=\vv_j-\vv_i$.
$\theta_{ij}$ is a symmetric random variable satisfying
\begin{eqnarray}\label{theta_equations}
\langle\theta_{ij}(t)\rangle&=&0,\\
\langle\theta_{ij}(t)\theta_{kl}(t')\rangle
&=&(\delta_{ik}\delta_{jl}+\delta_{il}\delta_{jk})\delta(t-t').
\end{eqnarray}
with $i\ne j$ and $k\ne l$. In Eq.(\ref{eq:random-force}), $\Delta t$ is the iteration time step. The weight factor is
chosen as
\BE\label{eq:omega}
\omega(r)=\left\{\begin{array}{ll}
1-r/r_c & \mbox{ for $r \leq r_c$,}\\
0 & \mbox{ for $r > r_c$}
\end{array}
\right.
\EE
where $r_c$ is the interactions cutoff radius. The choice of $\omega$ in Eq. (\ref{eq:omega})
ensures that the interactions are all soft and repulsive. 
The values of the parameter $a_{ij}$ used in the simulations are specifically chosen as
\BE
a_{ij}=\frac{\epsilon}{r_c} \left(\begin{array}{cccc}
   &w  &s  &p \\
w& -  &3  &10\\
s& 3   &25 &25\\
p& 10  &25 &50\\
\end{array}
\right),
\EE
where $w$, $s$ and $p$ designate a wall, solvent and a monomer particle, respectively.
We should not that $a_{ww}$ is irrelevant since the wall particles are kept frozen in the simulation.
The importance of the wall-solvent interaction to the velocity field boundary condition
will be discussed in Section III.
The integrity of a polymer chain is ensured via an additional harmonic interaction between consecutive monomers,
\BE\label{eq:spring-force}
{\bf F}_{i,i+1}^{(S)}=-C \left(1-r_{i,i+1}/b\right){\nn}_{i,i+1},
\EE
where we set, for the spring constant and the preferred bond length,
$C=100\epsilon$ and $b=0.45r_c$, respectively.

The equations of motion of particle $i$ are given by
\begin{eqnarray}\label{eq:motion}
d\rr_i(t)&=&\vv_i(t) dt,\\
d\vv_i(t)&=&\frac{1}{m}\sum_j [\FF^{(C)}_{ij} dt+\FF^{(D)}_{ij} dt \nonumber \\
         &+&\FF_{ij}^{(R)} (dt)^{1/2}],
\end{eqnarray}
where $m$ is the mass of a single DPD particle. Here, for simplicity, masses of all
types of dpd particles are supposed to be equal.
Assuming that the system is in a heat bath at a temperature $T$, the
parameters $\gamma_{ij}$ and $\sigma_{ij}$ in Eqs.~(\ref{eq:dissipative-force})
and (\ref{eq:random-force}) are related to each other by the fluctuation-dissipation theorem,
\BE \label{eq:fluctuation-dissipation}
\gamma_{ij}=\sigma_{ij}^2/2k_{\rm B}T.
\EE

Consider a fluid in a pore of cross section ${\cal A}$ under 
a pressure gradient ${\partial P}/{\partial x}$.
As a result, a portion of fluid of length $dx$ along the $x$-axis
experiences a net force along the direction of diminishing 
$P$ given by, $d{\bf F}={\cal A}dP \hat{\bf x}$. Consequently, each
particle $i$ in the volume ${\cal A}dx$ experiences an additional force 
\BE
{\bf f}_i= f_x \hat{\bf x}=-\frac{1}{\rho}\biggl{|}\frac{\partial P}{\partial x}\biggl{|}\hat{\bf x},
\EE
to be added to the forces in Eqs.~(1-3). The values of the driving force considered in the
present study range from 0 to $0.04\epsilon/r_c$. 

Most of our simulations are performed on boxes of dimensions $L_x\times L_y\times L_z$ with
$L_x=L_y=19.6r_c$ and $L_z=20r_c$.  We made sure that finite size effects are absent by performing
simulations on boxes with larger values of $L_x$ and $L_y$. 
As will be discussed in the next section, a frozen wall
is used. The wall is parallel to the $xy$-plane, and the flow is along the $x$-axis.

We used periodic boundary conditions in the three directions. Therefore only one wall is used.
We consider a fluid with a number density $\rho=4 r_c^{-3}$, $k_BT=\epsilon$, 
and $\sigma=3.0(\epsilon^3m/r_c^2)^{1/4}$. 
We used a wall with thickness $2.8r_c$. The wall is constructed from dpd particles that are arranged in 
a face centered cubic lattice with number density $61.35 r_c^{-3}$, and therefore much higher than that of the
solvent. This is done in order to prevent the solvent particles from penetrating the wall.
The iteration time $\delta t=0.01\tau$, where the time
scale $\tau=(mr_c^2/\epsilon)^{1/2}$. All our simulations were at least run over a time period
of $8000\tau$, corresponding to $8\times 10^5$ time steps. The first $1000\tau$ is used for relaxation of 
the system towards its steady state, and the remaining time is used for collection of the data. 

\section{POISEUILLE FLOW OF A SIMPLE FLUID AND PARAMETER SELECTION FOR WALL-SOLVENT INTERACTION}

Before presenting results on the flow of polymer solutions, it is useful to investigate the flow of 
the solvent alone (a Newtonian fluid) in a slit, using the DPD approach,
and compare with the theoretical expectations for Poiseuille flow. We recall that in a laminar
flow along the $x$-axis, the $xz$-component of the stress tensor,
$\Pi_{xz}$ is related to the shear rate, $(\partial v_x/\partial z)$, through the relation
$\Pi_{xz}=-\eta_0(\partial v_x/\partial z )$, where $\eta_0$ is the fluid viscosity.
The solution of the steady-state Navier-Stokes equation of the incompressible Newtonian fluid  
$\nabla P=\eta_0\nabla^2 {\bf v}$, with a no-slip boundary condition at the walls,
which are perpendicular to the $z$-axis, ${\bf v}\left(z=\pm H/2\right)=0$, 
and under a uniform pressure drop, $(dP/dx)$ along the $x$-axis,
corresponds to the usual Hagen-Poiseuille parabolic profile
\cite{landau}
\BE\label{eq:poiseuille}
v_x(z)=-\frac{H^2}{8\eta_0}\left(\frac{d P}{d x}\right)\left[1-\left(\frac{z}{H}\right)^2\right].
\EE

Within the DPD and molecular dynamics approaches, 
several techniques have been proposed to provide a no-slip wall-fluid boundary conditions. 
Revenga {\it et al.}, for example, used an effective field, 
instead of a wall composed of particles, in conjunction 
with a reflection force to reflect particles that cross the wall~\cite{revenga98}.
Willemsen {\etal.}~\cite{willemsen00} proposed a method that involves a layer of 
particles along the wall outside the 
simulation box. The coordinates and velocities of these particles are determined by mirroring those of the fluid particles 
within the cutoff distance of the interaction potential. The momenta of the additional particles are such that the average
velocity of a fluid particle, within the cutoff distance, and its auxilary particle within the wall satisfies 
the desired boundary condition.  
In a recent study, we found that the use of a thermalized wall can also help 
in obtaining a no-slip boundary condition~\cite{huang06}. 
In the current study, we found that a no-slip boundary condition can practically be obtained 
by lowering the wall-fluid interaction.

\begin{figure}
\includegraphics[scale=0.40]{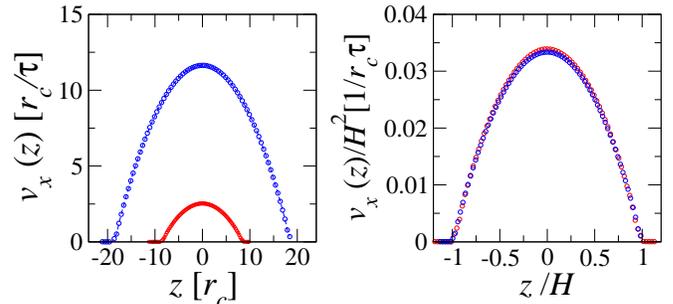}
\caption{Velocity field along the $x$-axis vs. $z$ for $a_{ws}=3\epsilon/r_c$.
Curves from bottom to top correspond to a driving force $f_x=0.01$, 0.02, 0.03, and $0.04\epsilon/r_c$.
Inset shows the same velocity profile normalized by the driving force. The excellent collapse of all
data is an implication that the velocity profile follows Eq.~(\ref{eq:poiseuille}).}
\label{fig3}
\end{figure}

In Fig.~\ref{fig1}, the velocity profile of a pure solvent is shown for the case of a slit width
$L_z=20 r_c$, a driving force $f_x=0.02\epsilon/r_c$ and values of the 
wall-solvent interaction corresponding to $a_{ws}= 3\epsilon/r_c$, $10\epsilon/r_c$, $20\epsilon/r_c$,
and $30\epsilon/r_c$. This figure shows that the velocity profile exhibits a discontinuity
when the amplitude is large ($a_{ws}> 10\epsilon/r_c$), implying that the fluid lubricates the wall 
({\em i.e.} the wall does not act as a no-slip boundary) at large wall-solvent interactions. However for 
$a_{ws}\leq 10\epsilon/r_c$, the wall acts effectively as a no-slip boundary condition. In the remaining if this
article, all simulations were performed with a wall-solvent interaction parameter $a_{ws}=3\epsilon/r_c$.

In Fig.~\ref{fig2}(a), the velocity profile of the simple fluid is shown for different values of the driving force, and
a wall solvent interaction, $a_{ws}=3\epsilon/r_c$. As shown in Fig.~\ref{fig2}(b), 
the normalization of the $z$-component of the
velocity field by the driving force yields a universal velocity profile, in excellent
agreement with the theoretical prediction for Poiseuille flow, Eq. (\ref{eq:poiseuille}).  
We also tested the scaling of the velocity profile
with the width of the slit pore. As shown in Fig.~\ref{fig3}, we found again that the scaling of the
velocity profile with the width of of the slit is in excellent agreement with Eq. (\ref{eq:poiseuille}).

\begin{figure}
\includegraphics[scale=0.35]{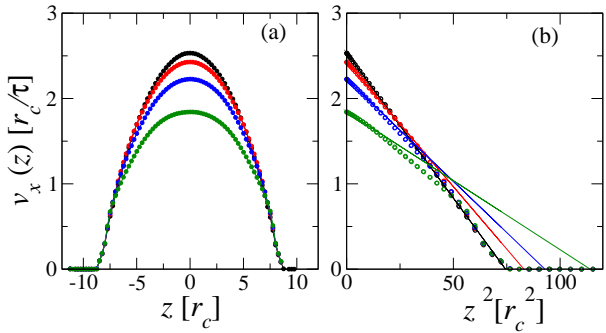}
\caption{(a) Velocity field along the $x$-axis vs. $z$ for a polymer solution
with $N=100$ and at a driving force $f_x=0.02\epsilon/r_c$.
Curves from top to bottom correspond to polymer volume fraction
$\varphi_p=0$, $0.06$, $0.12$ and $0.24$, respectively. $a_{ws}=3\epsilon/r_c$.
(b) Same data in (a) plotted vs. $z^2$. The straight lines are linear fits to
the data for small values of $z$.}
\label{fig4}
\end{figure}

\section{POISEUILLE FLOW OF POLYMER SOLUTIONS}

\subsection{Velocity Profile of Polymer Solutions in Poiseuille Flow}
For a polymer solution, which is a non-Newtonian fluid, the shear stress relation to the shear rate,
$(d v_x/d z)$,  
is approximated by a power law equation proposed by Ostwald~\cite{wilkinson60},
\BE
\Pi_{xz}=-K\left(\frac{d v_x}{d z}\right)^n,
\EE
where $K$ and the exponent $n$, are usually referred to as the flow consistency index and the fluid
power-law index, respectively. For a polymer solution, which is a shear-thinning fluid, 
the power-law index $n<1$. Eq.~(1) translates into a solution viscosity $\eta \sim (dv_x/dz)^{n-1}$. 
Note that Eq.~(1) implies that the viscosity diverges for very low shear rates. Therefore, the power
law model is not expected to hold at very low shear rates.  

The balance between the driving force due to pressure gradient and the forces due to shear stresses
lead to the solution,
\begin{eqnarray}\label{eq:poiseuille-non-newtonian}
v_x(z)&=&\frac{n}{1+n}K^{-1/n}\biggl{|}\frac{dP}{dx}\biggl{|}^{{1}/{n}}
H^{{1}/{n}+1} \nonumber \\
&\times&\left[1-\left(\frac{z}{H}\right)^{{1}/{n}+1}\right],
\end{eqnarray}
which reduces to Eq.~(\ref{eq:poiseuille}) when $n=1$.

\begin{figure}
\includegraphics[scale=0.47]{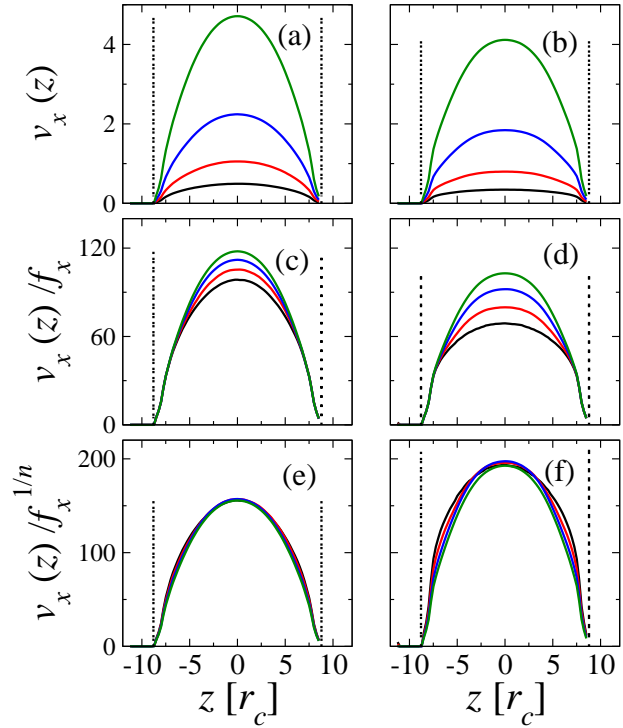}
\caption{Fluid velocity profile across the slit for the case of
(a) $\varphi=0.12$ and $N=50$ and (b) $\varphi=0.24$ and $N=100$.
In (a) and (b) Curves from bottom to top correspond
to $f_x=0.005$, $0.01$, $0.02$, and $0.04\epsilon/r_c$.
The velocity profiles scaled by the driving force are shown for (c) $\varphi=0.12$
and $N=50$, and (d) $\varphi=0.24$ and $N=100$. The velocity profiles scaled by $f_x^{1/n}$ are
shown in (e) $\varphi=0.12$ and $N=50$, and (f) $\varphi=0.24$ and $N=100$. In (e), $n=0.92$ and
in (f), $n=0.84$. The vertical dot-dashed lines indicate the positions of the walls.}
\label{fig5}
\end{figure}

We performed a large number of systematic simulations of flow of polymer solutions for different values of
chain length corresponding to $N=5$, 20, 50, and 100, polymer volume fraction corresponding to
$\varphi=0.06$, 0.12, and 0.24, and driving force ranging between 0 and $0.04\epsilon/r_c$.
In Fig.~\ref{fig4}, the velocity profile of a polymer solution with chain length $N=100$ and under a driving
force $f_x=0.02\epsilon/r_c$ is shown for volume fractions, $\varphi=0$, $0.06$, $0.12$ and $0.24$. This
figure shows that except from the region close to the wall, the velocity profile is not quadratic in
$z$, and therefore does not follow Hagen-Poiseuille's law for Newtonian fluids, 
Eq. (\ref{eq:poiseuille}). This is due to the non-linear
rheological properties of the polymer solution.
In vicinity of the wall, however, the velocity profile is practically independent of $\varphi$. This
is presumably due to the fact that the polymer chains are sterically depleted from the wall regions.
In Fig.~\ref{fig4}(b), an approximate linear fit of the velocity profile vs. $z^2$ around the mid-section of the slit
leads to an effective wall boundary that extends beyond the physical boundary. This leads to the an apparent slip
of the polymer solution at the wall.

In order to investigate even further the departure of the velocity profile from that Hagen-Poiseuille quadratic 
profile of Newtonian fluids, the velocity profiles of two polymer solutions, corresponding to $(\varphi,N)=(0.12, 50)$
and $(0.24,100)$ are shown in Fig.~\ref{fig5}.  Fig.~\ref{fig5}(c) and (d) show that,
in contrast to the case of a simple fluid  (see Fig.~\ref{fig2}(b)), 
the velocity profile of a polymer solution does not scale 
linearly with the driving force, and that the departure from this scaling worsens as the chain length 
and/or the volume fraction of the polymers is increased. It is interesting to note that the scaling holds reasonably well
close to the walls. This is due to the very low polymer density near the walls.
A fit of the velocity profile with the generalized
Hagen-Poiseuille equation for Non-Newtonian fluids, 
Eq.~(\ref{eq:poiseuille-non-newtonian}), yields  power
law fluid index $n=0.92$ and $n=0.84$ for the case of $(\varphi,N)=(0.12, 50)$ and $(0.24,100)$, respectively. 
A modified scaling of the velocity profile with $f_x^{1/n}$ is depicted 
in Figs.~\ref{fig5}(e) and 5(f) respectively. Fig.~\ref{fig5}(e) 
shows that the scaling holds reasonable well in the case of relatively low volume fraction of
polymer. Fig.~\ref{fig5}(f) shows that the scaling does not hold as well 
in the case of $(\varphi,N)=(0.24, 100)$, where both
the polymer volume fraction and molecular weight are higher.
This is presumably due to the fact that the exponent $n$ itself 
depends on the driving force, $f_x$. 

\begin{figure}
\includegraphics[scale=0.37]{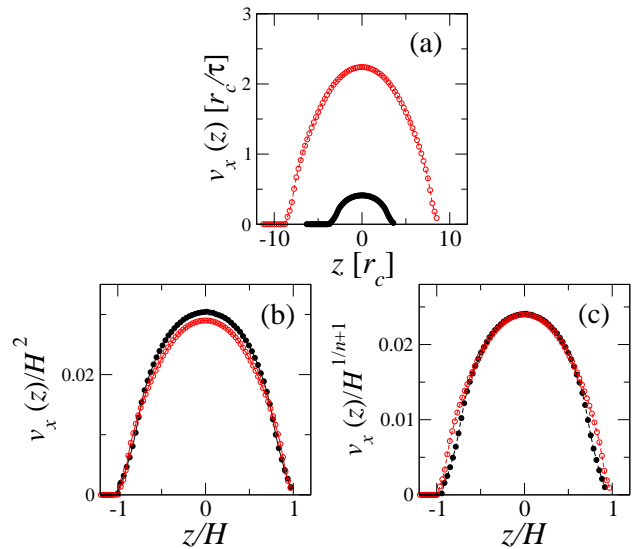}
\caption{(a) Fluid velocity profile across the slit for the case of $\varphi=0.12$,
$N=50$ and $f_x=0.02\epsilon/r_c$. Solid circles and open circles correspond to a 
slit thickness $H=7.8 r_c$ and $H=17.6 r_c$, respectively. 
(b) The fluid velocity scaled by $H^2$ vs. $z/H$. 
(c) The fluid velocity scaled by $H^{\frac{1}{n}+1}$ vs. $z/H$,  with $n=0.84$.}
\label{fig6}
\end{figure}

In Fig.~\ref{fig6}, the velocity profile for a solution with $(\varphi,N)=(0.12, 50)$
under a driving force $f_x=0.02\epsilon/r_c$ is shown for two values of the slit thickness
corresponding to $H=7.8r_c$ and $17.6r_c$. Fig.~\ref{fig6}(b) shows that the scaling of the velocity 
profile with $H^2$ in the case of a polymer solution does not hold in the center-plane region, 
but holds reasonably well in vicinity of the walls. This is again due to the depletion of polymers 
from the walls region.
Fig.~\ref{fig6}(c) shows that a better scaling in the mid-section region is achieved 
when Eq.~(\ref{eq:poiseuille-non-newtonian}) is used with $n=0.84$.  

We now turn our attention to the analysis of the polymers velocity profile. 
Since the polymers are sterically 
depleted from the regions near the wall, it is expected that in average the polymer chains move faster 
than the solvent and that the longer polymer chains move faster than the shorter 
polymer chains~\cite{dimarzio70}.  In Fig.~7, the ratio between the average polymer velocity and
the average fluid velocity, $\langle v^p_x\rangle/\langle v_x\rangle$ is shown. 
This figure shows that the polymer moves faster than the solution
by about $10\%$. This effect is however reduced as the polymer volume fraction is increased. This is due 
to the fact that across the slit, the polymers become more evenly distributed as their volume fraction
is increased. Fig.~7 shows as well that as the chain length is increased, 
$\langle v^p_x\rangle/\langle v_x\rangle$ increases, mainly as a result of exclusion of the 
polymers from the walls. However, for $N=100$, Fig.~7 shows that $\langle v^p_x\rangle/\langle v_x\rangle$ 
is slightly smaller than that for shorter chains. Presumably, this is due to the fact that long chains 
tend to migrate away from the center-plane, as will be discuss later.

\begin{figure}
\includegraphics[scale=0.37]{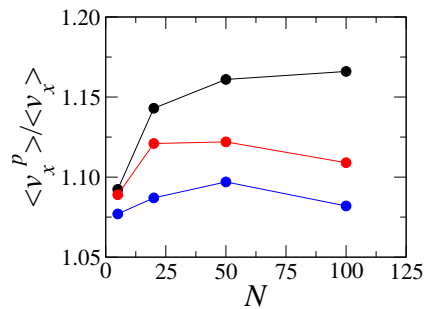}
\caption{The ratio between average polymer velocity and solution velocity along the flow direction,
$\langle v^p_x\rangle/\langle v_x\rangle$ as a function for chain length for $f_x=0.02\epsilon/r_c$ and
$H=7.8r_c$.
Curves from top to bottom correspond to $\phi=0.06$, $0.12$ and $0.24$, respectively.}
\label{fig7}
\end{figure}

\begin{figure}
\includegraphics[scale=0.35]{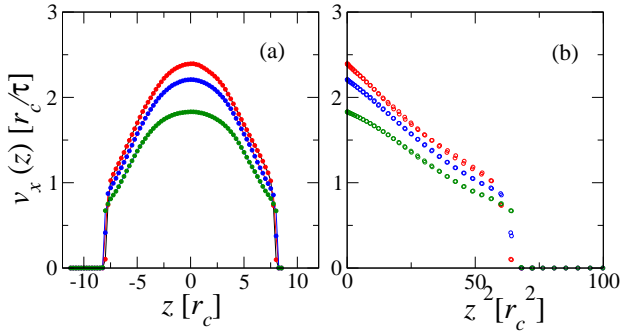}
\caption{(a) Polymer velocity field along the $x$-axis vs. $z$
for a polymer solution with $N=100$ and at a driving force
$f_x=0.02\epsilon/r_c$.
Curves from top to bottom correspond to polymer volume fraction
$\varphi_p=0.06$, $0.12$ and $0.24$, respectively. $a_{ws}=3\epsilon/r_c$.
(b) Same data in (a) plotted vs. $z^2$.  }
\label{fig8}
\end{figure}

In Fig.~\ref{fig8}, the velocity profile of the polymer chains is shown for the case of $N=100$ at a driving force 
$f_x=0.02\epsilon/r_c$ for volume fractions $\varphi=0.06$, 0.12 and 0.24. 
This figure shows that the velocity profile
of the polymer chains differs from that of the embedding solvent despite the fact that the polymer chains are 
transported by the solvent. Fig.~\ref{fig9}(a) and (b) depict
the velocity profiles of the fluid and the polymer for
the case of $(\varphi,N)=(0.06,5)$ and $(0.06,100)$, respectively.
There, it is clear
that the two profiles do not coincide, particularly when the chain length or volume fraction is increased. 
Fig.~\ref{fig8} shows that the polymer velocity 
exhibits a sharp discontinuity near the walls of the slit, implying that as far as the polymer is concerned, the
walls act as a slip boundary. Fig.~\ref{fig9} shows in the middle of the slit, the polymer chains move slightly slower
than the embedding solvent, and that the difference between the solvent velocity and the polymer
velocity is increased as the polymer chain length or volume faction is increased. Fig.~\ref{fig9} also shows that in the
middle of the slit the polymer motion is retarded as compared to that of the solvent. However, the motion 
the polymers is faster than the solvent near the walls. The slower motion of the polymer chains near the walls
is due to the fact that due to the monomers connectivity, monomers belonging to a single chain have to move
with same average velocity.

\subsection{Polymers Conformational Distribution}

\begin{figure}
\includegraphics[scale=0.37]{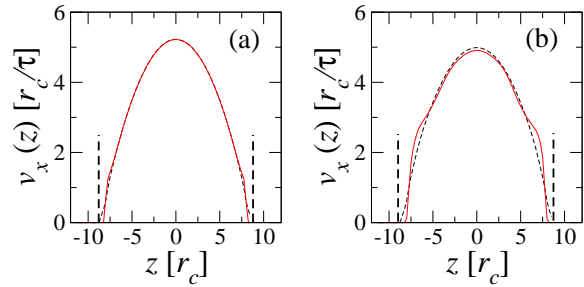}                                  
\caption{Solvent velocity profile (dashed line) is shown together with the polymer velocity profile 
(solid line) for the case of $\varphi=0.06$ and $f_x=0.04\epsilon/r_c$. 
(a) and (b) correspond to $N=5$ and $N=100$, respectively.
The vertical dot-dashed lines indicate the positions of the walls.}
\label{fig9}                                                            
\end{figure}

\begin{figure}
\includegraphics[scale=0.55]{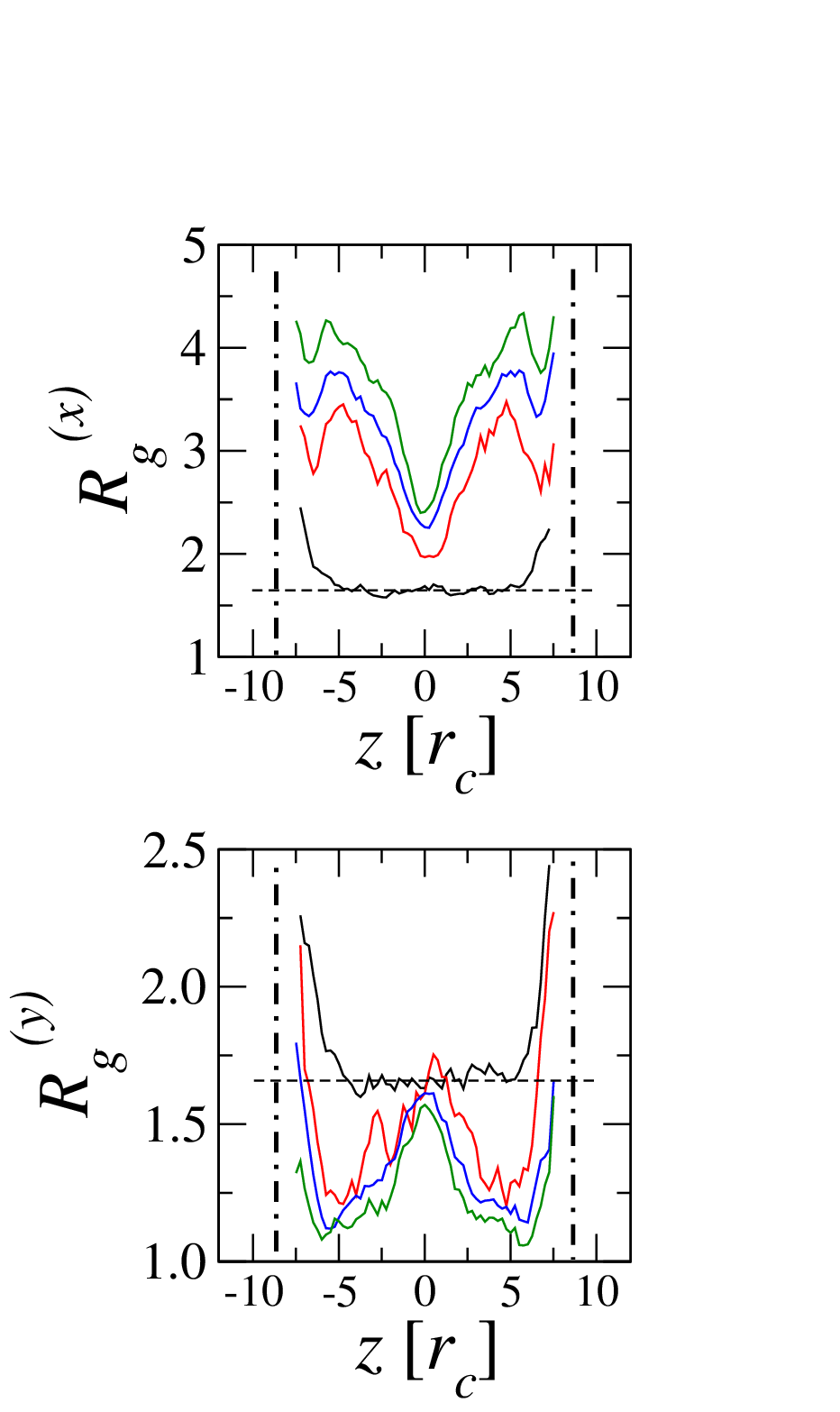} 
\includegraphics[scale=0.55]{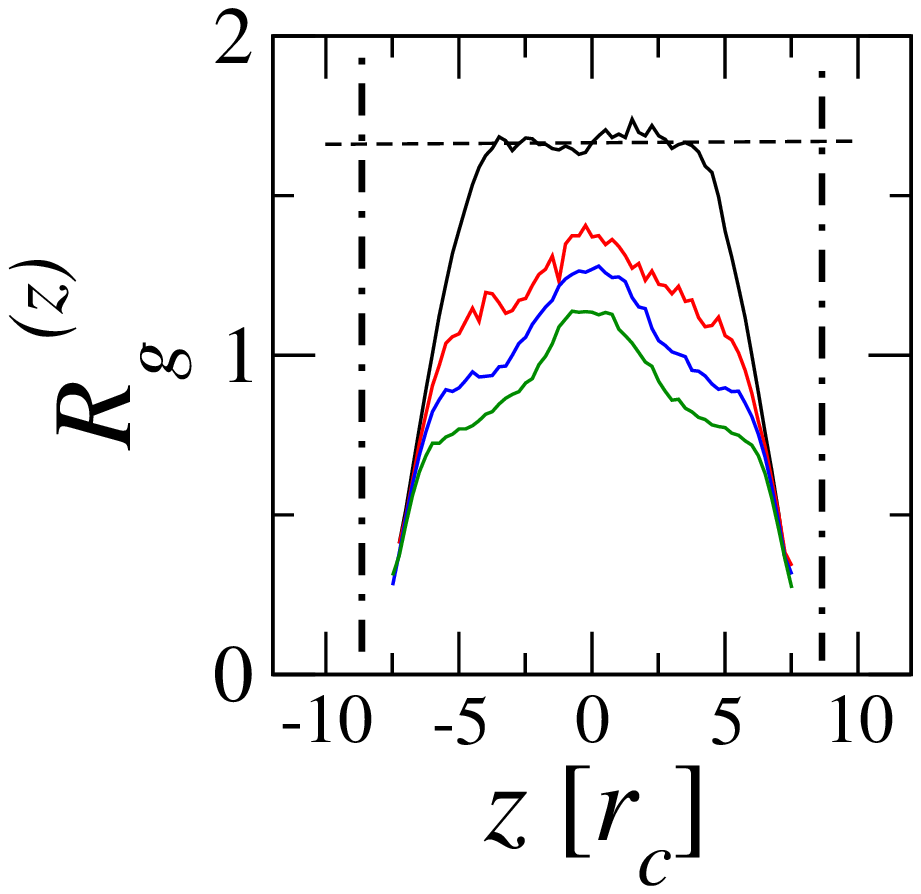} 
\caption{Distribution of the three components of chains radius of gyration along 
the slit for the case of $N=50$ and 
$\varphi=0.12$. In (x), curves from bottom to top correspond to
$f_x=0$, $0.01\epsilon/r_c$, $0.02\epsilon/r_c$ and $0.04\epsilon/r_c$. In (y) and
(z), curves from top to bottom correspond to
$f_x=0$, $0.01\epsilon/r_c$, $0.02\epsilon/r_c$ and $0.04\epsilon/r_c$.
The horizontal dashed line corresponds to the value of $R_g/\sqrt{3}$ of a bulk solution
at equilibrium. The vertical dot-dashed lines indicate the positions of the walls.}
\label{fig10}                                                                                 
\end{figure}

We now focus on the effect of flow on the conformational distribution of the polymer chains in solution.
In Fig.~\ref{fig10}, the distributions of the three components of the radius of gyration,
$R_g^{(x)}$, $R_g^{(y)}$ and $R_g^{(z)}$, of a polymer
solution, with $N=50$ and $\varphi=0.12$ are shown for driving forces corresponding
to $f_x=0$, $0.01\epsilon/r_c$, $0.02\epsilon/r_c$, and $0.04\epsilon/r_c$. The $z$-axis represents
the position of the center of mass of the chains.
This figure indicates that, near the walls, the $z$-component of the radius of gyration is lower than 
than $R_g^{(x)}$ and $R_g^{(y)}$ near the walls. This is due to the steric interactions with the walls.
Fig.~\ref{fig10} shows that at equilibrium ($f_x=0$) and away from the walls, 
the three components of the radius of gyration are equal, implying that the chains away from the walls
do not experience confinement. This implies that under non-equilibrium conditions,
the redistribution of the radius of gyration is due to the flow. 
As the driving force is increased, 
the chains become stretched along the flow direction, {\it i.e.} along the $x$-axis,
regardless of the location of their center of mass across the channel. 
This figure also shows that the chains stretching along the flow direction is lowest in the mid-section 
of the slit pore.
This is due to the fact that the shear rate, and therefore the resulting stresses, 
are lowest in this region.
Away from the mid-section of the slit, the chains stretching increases but in a non-monotonous fashion. 
In particular, we observe that the distribution of $R_g^{(x)}$ exhibits two peaks somewhat close, but not very close,
to the wall. The positions of these two peaks along the
$z$-axis approach the wall as the driving force is increased. 
Very close to the walls, and within the depletion layer, the stretching is again increased. 
A similar complex behavior of the distribution of chains radius of gyration was 
recently reported by Jendrejack {\it et al.} using their self-consistent Langevin dynamics 
approach~\cite{jendrejack04}. We also note that the position of the two weak off-center 
maxima in the distribution of the $x$-component correlate with the onset of the shoulders
in the velocity profile of the polymer chains, shown in Fig.~\ref{fig9}(b). 

Fig.~\ref{fig10} shows that the $y$-component of the radius of gyration,
$R_g^{(y)}$, decreases from its maximum value at the center-plane up to some point, and then increases at
distances closer to the wall. Since the shear stress $\Pi_{xy}=0$, the contraction
of the chains along the $y$-axis and around the center-plane is a partial compensation to the
chains stretching along the $x$-axis. The increase of the $y$-component of the radius of gyration, 
closer to the walls, is a compensation to the decrease of the $z$-component of the radius of gyration close 
to the walls.  As for the $z$-component, $R_g^{(z)}$, its fast decrease near the walls is due to steric interactions 
with the walls. around the mid-plane, the decrease is slower, and is due to gradient in shear rate along the $x$-axis. 
As the driving force is increased, shear stresses are amplified leading to further contraction of the chains along
the $z$-axis.

\begin{figure}
\includegraphics[scale=0.55]{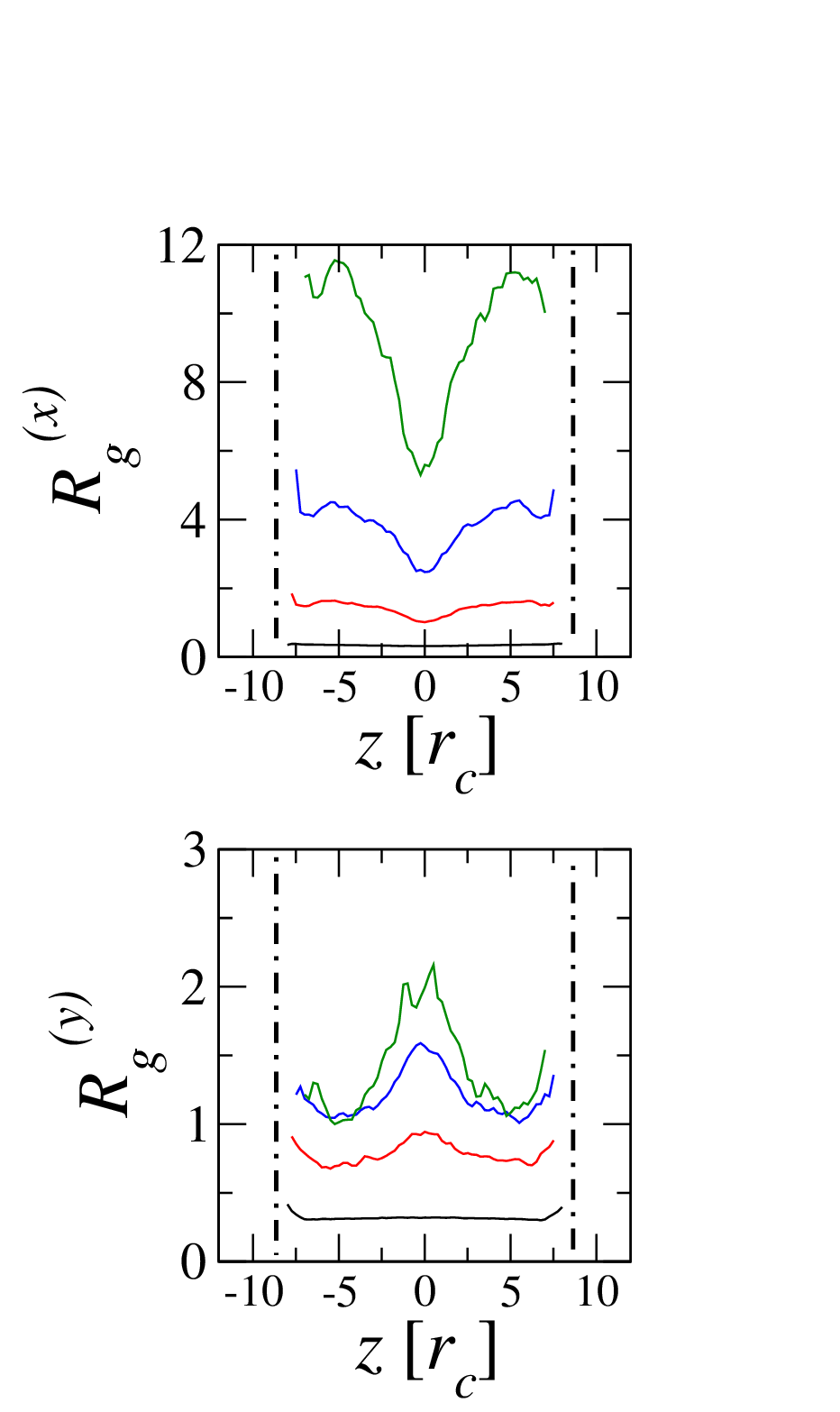}                                      
\includegraphics[scale=0.55]{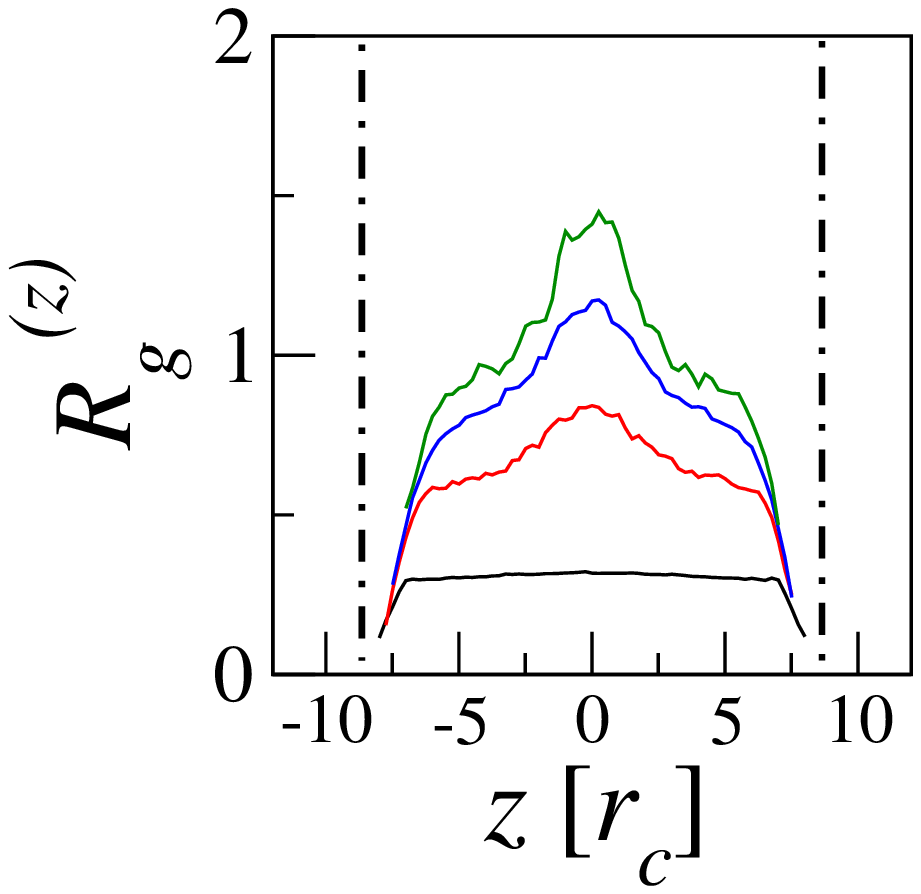}                                      
\caption{Distribution of the three components of chains radius of gyration along 
the slit for the case of $f_x=0.04\epsilon/r_c$ and $\varphi=0.12$.
Curves from bottom to top correspond to $N=5$, $20$, $50$, and $100$, respectively.
The vertical dot-dashed lines indicate the positions of the walls.}
\label{fig11}
\end{figure}

In Fig.~\ref{fig11}, the profiles of the components of the radius of gyration 
across the channel are shown for a solution
with $\varphi_p=0.12$ with $N=5$, 20, 50, and 100 at a driving force $f_x=0.04\epsilon/r_c$. This figure shows that
the effects observed in Fig.~\ref{fig10} are enhanced as the chain length is increased. 

\subsection{Polymers Density Profile}
\begin{figure}
\includegraphics[scale=0.4]{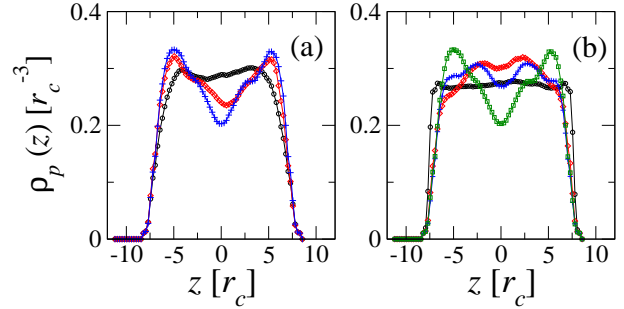}
\caption{(a) Polymer density profile across the slit for the case of 
$\varphi=0.06$ and $N=100$ for varying values of the driving force: 
$f_x=0$ (squares), $f_x=0.02\epsilon/r_c$ (diamonds), and 
$f_x=0.04\epsilon/r_c$ (pluses). 
(b) Polymer density profile for the case of $\varphi=0.06$, 
$f_x=0.04\epsilon/r_c$, and chain length $N=5$ (circles), $N=20$ (diamonds),
$N=50$ (pluses), and $N=100$ (squares).}
\label{fig12}
\end{figure}

In Fig.~\ref{fig12}(a), the monomers distribution in the case of a dilute
solution with $\varphi=0.06$ and $N=100$ is shown for
three values of the driving force corresponding to 
$f_x=0$, 0.02 and $0.04\epsilon/r_c$. 
This figure shows that at equilibrium conditions
($f_x=0$) and away from the walls, the distribution is, as expected, essentially uniform. However, 
in the presence of flow, the distribution 
exhibits a local minimum in the center-plane.  
This figure also shows that the thickness of the depletion layer slightly decreases 
as the driving force is increased. This figure
therefore implies that for this specific $\varphi$ and $N$, the polymer chains migrate away 
from the center-plane towards the walls. We note that in this case the ratio between the slit thickness 
and the chain radius of gyration is $2H/R_g\approx 3.6$. For this ratio, Usta {\it et al.} also found
a migration towards the wall~\cite{usta06}. In the case of weak confinement, $\phi=0.06$ and $N=20$ 
(for which $2H/R_g\approx 10.8$), we
also found a peak in the middle of the slit, although the peak is weaker than for $N=100$. Near the walls, we found
a slight increase in the depletion layer, again in qualitative agreement with Usta {\it et al.}'s
work~\cite{usta06}. The presence of a dip, two off-center peaks and the slight deviation in the depletion layer
near the walls was also observed in the recent molecular dynamics simulations by Khare {\it et al.}
~\cite{khare06}.

In Fig.~\ref{fig12}(b), the monomers distributions for $\varphi=0.06$, and chain length $N=5$, 20, 50
and 100 are shown for a driving force $f_x=0.02\epsilon/r_c$. This figure shows, that the while the distribution is flat for small chains ($N=5$), a migration towards the center-plane is clearly observed as
the chain length is increased ($N=20$), as demonstrated by an increase in the thickness
of the depletion layers near the walls. However, at the very center-plane, the density profile exhibits
a local minimum. As the chain length is further increased, the density profile in the center-plane
is further decreased, while the two off-center peaks are increased. the depletion layer near the walls 
ultimately narrows, but slightly, as the chain length is increased. This rich behavior of the polymer density profile
is qualitatively in agreement with that reported by Usta {\it et al.} using a lattice Boltzmann simulation
~\cite{usta06}, although in their case the size of the depletion layer in the weak confinement regime seem
to be strongly affected by the flow rate. 

The presence of a dip in the center-plane of the slit in Poiseuille flow implies a migration away
from the very center of the slit. However, this can be accompanied by an increase in the depletion
layer at the walls. Therefore an overall migration of the polymers cannot be merely inferred by merely
looking at the depletion layer. We propose that the cross-stream migration issue
can better be investigated through the second moment of the polymer volume fraction profile,
\BE
\langle z^2\rangle={\int_{-H}^H dz z^2\rho_p(z)}/{\int_{-H}^H dz\rho_p(z)}.
\EE
This is shown as a function of chain length, $N$,
for all systems investigated in Fig.~13. A small $\langle z^2\rangle$ implies a
general migration from the walls. This figure shows that in the absence of flow ($f_x=0$), 
$\langle z^2\rangle$ decreases as $N$ increases. This is due to the increase in the depletion layer as
$N$ is increased. In the presence of flow, $\langle z^2\rangle$ decreases with increasing $N$
for small values of $\varphi$. However, for relatively large values of polymer volume fraction, 
$\langle z^2\rangle$ decreases with increasing $N$ for small values of $N$, but for large values of 
$N$, $\langle z^2\rangle$ increases with increasing $N$. Furthermore, Fig.~13 shows that for small
values of $N$, the level of migration increases with increasing the driving force, $f_x$. However, 
for large values of $N$, more migration towards the wall is observed as $f_x$ is increased.

\begin{figure}
\includegraphics[scale=0.5]{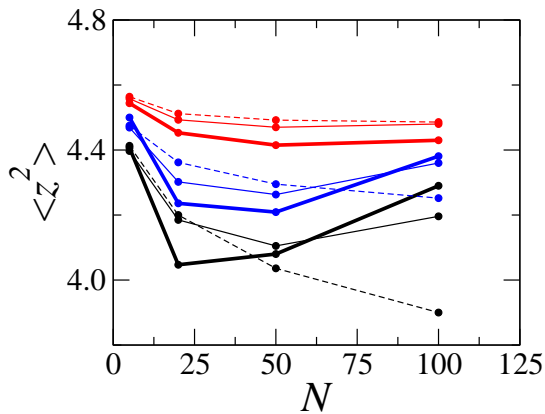}
\caption{The second moment as calculated from Eq.~(16) vs. chain length.
Dashed lines correspond to $f_x=0$, thin solid lines correspond to
$f_x=0.02\epsilon/r_c$,and thick solid lines correspond 
to $f_x=0.04\epsilon/r_c$. Black, blue and red lines correspond to 
$\varphi=0.06$, 0.12, and 0.24, respectively.}
\label{fig13}
\end{figure}

\section{Summary and Conclusions}

We presented results obtained from a systematic computational study of flow of polymer solutions in nanoscale slit
pores under the action of pressure gradient, using the dissipative particle dynamics approach. We found that
while the flow of a simple fluid  follows the Poiseuille-Hagen law, 
the flow of a polymer solution is characterized by a non-quadratic
velocity profile and a lack of scaling with respect to both driving force and slit thickness that fits relatively
well the generalized Poiseuille-Hagen law for non-Newtonian fluids with a power less than one. We also found that
while the polymer chains are extended along the channel,
the distribution of the three components of the radius of gyration is highly non-uniform across the channels.

The density profile of the polymers in the presence of pressure driven flow is also found to be non-uniform, exhibiting 
a local minimum in the center-plane and two off-center symmetric peaks. We found either a migration from or towards the 
wall depending on chain length, volume fraction and driving force. For relatively short chains and small
polymer volume fractions, a migration away from the walls is observed. 
In this case, the level of migration is amplified as the chain length or driving
force is increased. However, for longer chains a migration towards the wall is observed with the level of migration
is increased as the chain length or the driving force is increased. For relatively concentrated polymer solutions,
the level of migration is decreased. 
 
It was argued earlier by Groot and Warren~\cite{groot97} 
that DPD suffers from a low Schmidt number, ${\rm Sc}=\eta/\rho D_0\approx 1$, 
where $\eta$ is the viscosity, $\rho$ is the fluid density, and $D_0$ is the diffusivity of the solvent particles. 
In contrast a typical fluid has a Schmidt number of order 1000. A low  Schmidt number implies that the particles
diffuse as fast as momentum. Consequently, such an argument would lead us to conclude that the dynamics of polymer chains
from DPD in a dilute solution will not follow the Zimm model. However, a recent extensive DPD study by us has shown that 
polymer chains in dilute solution correctly obey the Zimm model. Therefore, within DPD, the hydrodynamic interaction 
is well developed within the time scale of polymer motion. As noted recently by Peters~\cite{peters04}, 
the Schmidt number in a 
coarse-grained model is in fact ill-defined, since in the Schmidt number, $D_0$ corresponds to the diffusion coefficient of
single particle, not coarse-grained fluid elements. Our current findings provide
further testimony to the effectiveness of the dissipative particle dynamics approach in 
simulating polymer flow in a variety of situations including in nano-channels. 
 
\section*{Acknowledgments}
This work was supported by a grant from the Petroleum Research Fund 
and a grant from The University of Memphis Faculty Research Grant Fund. The latter support does not necessarily imply 
the endorsement by the University of research conclusions. 
MEMPHYS is supported by the Danish National Research Foundation.

\end{document}